# Laboratoire de l'Accélérateur Linéaire

# Prospects for and status of the measurement of high x parton distributions at HERA

Zhiqing ZHANG

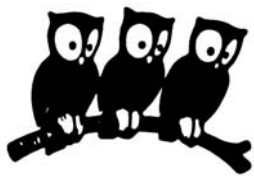

# Prospects for and Status of the Measurement of High x Parton Distributions at HERA

Zhiqing Zhang

Laboratoire de l'Accélérateur Linéaire
IN2P3-CNRS et Université de Paris-Sud, Bât. 200, BP 34, 91898 Orsay Cedex, France

**Abstract.** Inclusive neutral and charged current cross sections and structure functions measured at HERA-I are briefly reviewed. The impact of these measurements on parton density functions is discussed. The emphasis is put on the high x region. A first result obtained using the HERA-II data is also shown followed by future prospects for HERA.

## 1 Introduction

The HERA electron- or positron-proton collider is a unique facility for studying neutral and charged current (NC, CC) processes in deep inelastic scattering (DIS) interactions in an extended kinematic region compared to that covered previously by fixed target experiments. HERA-I operation started in 1992 and ended in 2000. Earlier measurements from H1 and ZEUS revealed the strong rise of the structure function (SF) $F_2$ towards small x. The more precise measurement of NC and CC cross sections is based on three recent data samples of $e^+p$ taken in years from 1994 to 1997 at a center-of-mass (CM) energy of 301 GeV, $e^-p$ in 1998-1999 and $e^+p$ in 1999-2000 at 319 GeV. The corresponding integrated luminosities per experiment of the three data samples are around 40 pb$^{-1}$, 15 pb$^{-1}$ and 65 pb$^{-1}$, respectively. The HERA-I data have driven the determination of parton density functions (PDFs) in various global fits for a decade. The universal nature of the PDFs makes precise determination essential in order to provide reliable predictions for Standard Model (SM) processes and for searches for new physics at future hadron-hadron colliders such as the LHC. Since 2002, HERA has entered its second phase of operation (HERA-II) and is being run in higher luminosity mode providing in addition a longitudinally polarized positron beam for H1 and ZEUS. The improved luminosity is due to a finer beam focusing around the interaction points and to larger beam currents.

In the following, I shall first briefly review the precise cross section and SF data measured at HERA-I and their impact on the determination of PDFs and then show a first measurement made using the HERA-II data before discussing future prospects. I will not discuss here other interesting and important physics topics at HERA such as heavy flavor physics, diffraction and searches for new physics.

## 2 HERA-I achievements

Both NC and CC processes can be produced at HERA. The NC interaction is dominated by γ exchange in the t channel at low $Q^2$ and receives additional contributions from Z exchange and γZ interference when $Q^2$ values are comparable with the Z mass squared. The CC interaction takes

place by W exchange only. The NC cross section may be expressed in terms of three SFs as

$$\frac{d^2\sigma_{NC}(e^\pm p)}{dxdQ^2} = \frac{2\pi\alpha^2}{xQ^4}\left[Y_+\widetilde{F}_2 - y^2\widetilde{F}_L \mp Y_- x\widetilde{F}_3\right] \quad (1)$$

where $Y_\pm = 1\pm(1-y)^2$. The CC cross section may be expressed in a similar way.

The dominant SF $\widetilde{F}_2$ has been determined from the measured NC cross sections over more than 4 orders of magnitude both in x and $Q^2$ with an accuracy down to 2% (Fig.1) [1,2]. The contribution of the SF $\widetilde{F}_L$ is sizable only at high y and can only be determined indirectly for the moment [3]. The SF $x F_3$ is non-negligible at high $Q^2$ and is determined [1,2] using both $e^+p$ and $e^-p$ data samples with so far limited precision.

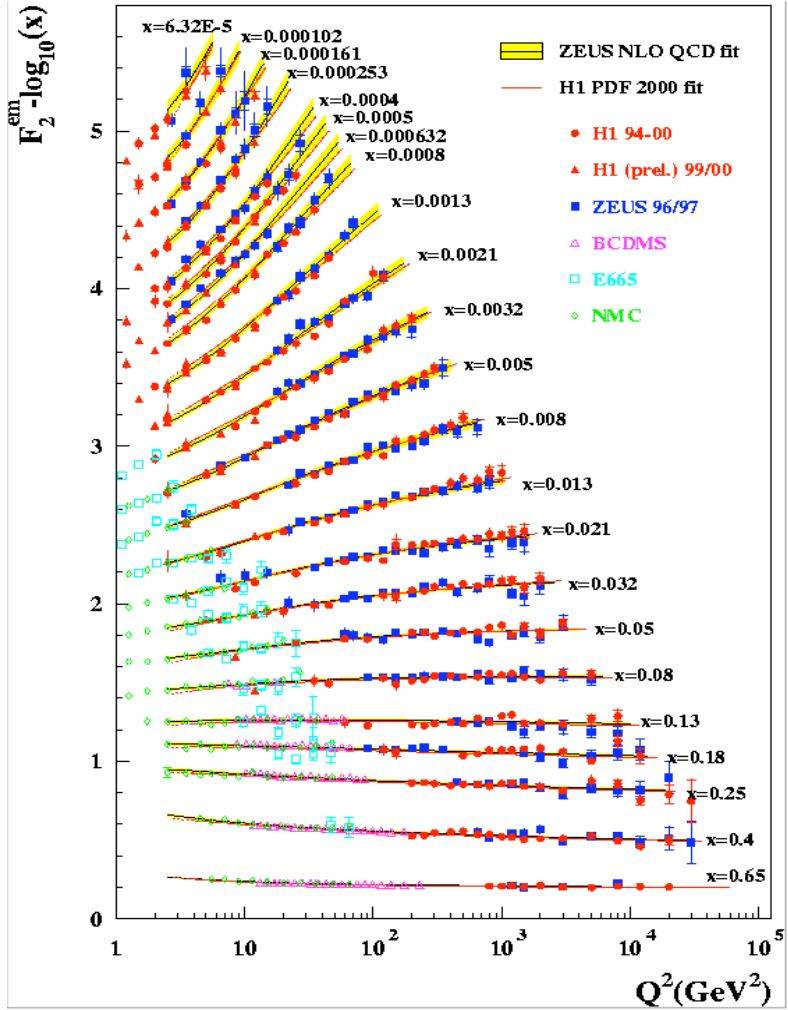

**FIGURE 1.** *The pure electromagnetic ($\gamma$ exchange) structure function $F_2^{em}$ in $\widetilde{F}_2$ measured by HERA experiments H1 and ZEUS and by fixed target experiments BCDMS, E665 and NMC. The curves show the corresponding expectations from ZEUS next-leading-order (NLO) QCD fit and H1 PDF 2000 fit (see text below).*

While the NC SF $\widetilde{F}_2$ is essentially flavor blind, the CC cross-sections are flavor dependent. In particular the measured CC $e^+p$ cross-section data [1,2] can be used to uniquely constrain the down type quark density (mainly d quark) free from any nuclear corrections.

HERA data have been extensively used by the MRST and CTEQ groups [4,5] in their global fits to determine the PDFs in the last decade with steadily increasing precision. The impact of the HERA-I data is best seen in QCD analyses using these data only. Indeed, the fit performed by H1 [1], based on all $e^+p$ and $e^-p$ NC and CC cross section data at low and high $Q^2$, was used to determine 5 PDF sets xP: xg, xU=x(u+c), xD=x(d+s), $x\overline{U}=x(\overline{u}+\overline{c})$ and $x\overline{D}=x(d+\overline{s})$. The PDFs were parameterized with the functional form $xP(x)=A_p x^{B_p}(1-x)^{C_p}[1+D_p x+E_p x^2+F_p x^3+G_p x^4]$ at $Q_0^2=4\text{GeV}^2$. The results of the fit are shown in Fig.2 compared to PDFs determined in a similar analysis from ZEUS [6] using, in addition to cross section data, jet data in order to improve the constraint on the gluon. The agreement between H1 and ZEUS is fairly good and the visible difference in e.g. the gluon distribution may be understood due to different functional forms, heavy flavor treatments and consistency of the data sets.

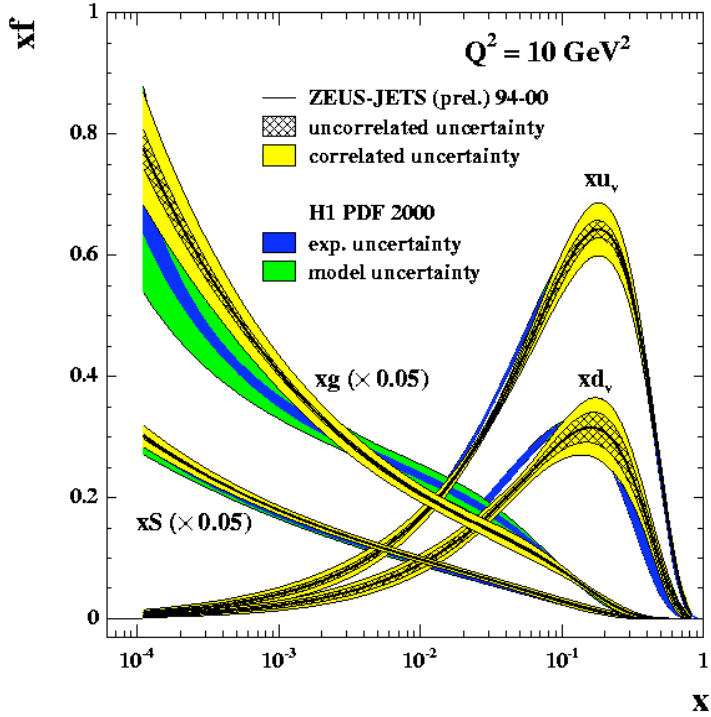

*FIGURE 2. A comparison of PDFs determined by H1 (H1 PDF 2000) based on inclusive NC and CC cross section data and by ZEUS (ZEUS-JETS) using jet data in addition to the cross section data. The gluon (xg) and sea (xS) densities are scaled by a factor 20.*

Another way [1] to see the constraining power of the HERA cross section data on the PDFs is to extract locally those parton distributions which contribute more than 70% in the relevant cross sections as:

$$xq(x,Q^2)=\sigma(x,Q^2)\left(\frac{xq(x,Q^2)}{\sigma(x,Q^2)}\right)_{\text{fit}}. \qquad (2)$$

The resulting xu and xd densities are shown in Fig.3. The determination is shown as a function of x for two values of $Q^2$ at 3000 and 8000 GeV$^2$ on the upper figures and as a function of $Q^2$ on the lower figures for x values above 0.08, the high x region where the method applies. Such a determination also compares well with the H1 PDF 2000 fit shown as the shaded bands and with MRST and CTEQ as curves.

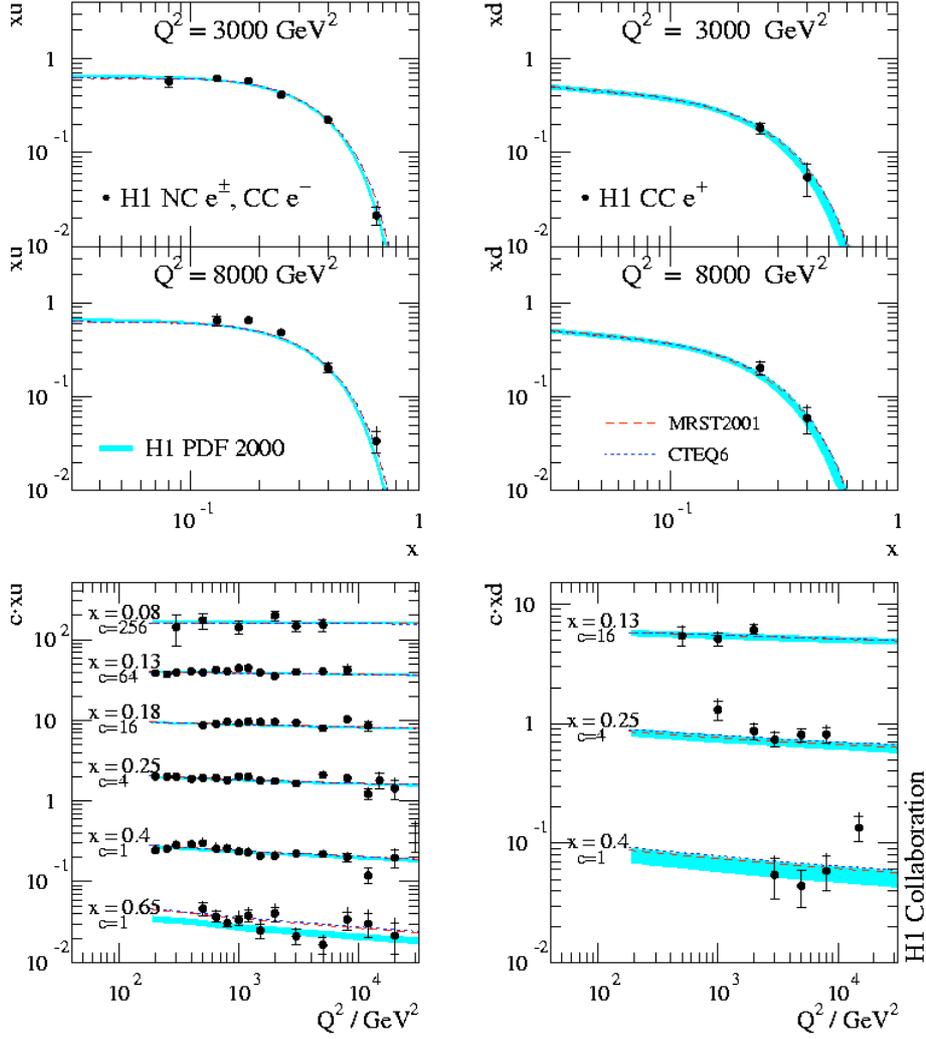

**FIGURE 3.** *The parton densities xu and xd determined in the local extraction method (data point) in comparison with those obtained from the H1 PDF 2000 fit (shaded band) and MRST and CTEQ6 (curve).*

## 3 HERA-II status and future prospects

In order to increase the luminosity delivered, the HERA accelerator was upgraded in 2001. During the first two years running after the upgrade however, the experiments suffered large backgrounds. The problem has now been solved and an increase in peak luminosity around a factor 3 was achieved in the running of 2004. In addition, two pairs of spin rotators have been installed near the H1 and ZEUS interaction points providing for the first time longitudinally polarized positron beams to these experiments.

HERA-II has already delivered two independent data samples of ~20 pb$^{-1}$ each with a mean polarization value of −44% and 33%, respectively. They have been used by both experiments to measure CC total cross sections. Together with the CC cross section measured [1,2] previously at HERA-I with unpolarized data, the dependence of the CC cross section on polarization is now available for the first time (Fig.4). In the SM, the cross section has a linear dependence as a function of polarization as represented by the full curve based on the expectation from MRST. Due to the absence of right-handed W bosons in the SM, the CC cross-section at $P_e=-1$ is predicted to vanish. Indeed, a straight line fit to the measured cross sections resulted in an extrapolated cross section value at $P_e=-1$ of $(-0.2\pm1.8[\text{stat}]\pm1.6[\text{syst}])$ pb, which is consistent with a null contribution. The result of the fit is shown as a dashed line in Fig.4. Future measurements with improved precision will provide a stringent constraint on the contribution of the right-handed W boson and its mass, independent of that from the Tevatron.

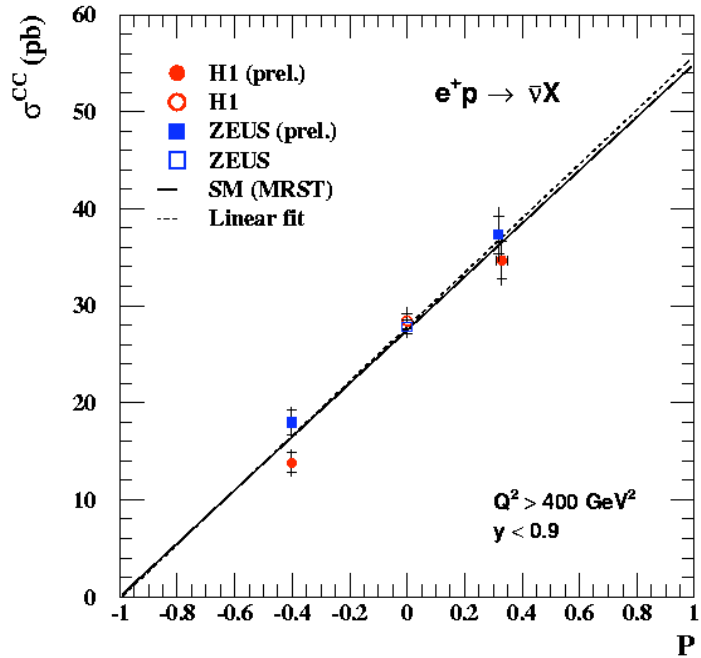

**FIGURE 4.** *The dependence on polarization of CC total cross sections shown for $Q^2>400$ GeV$^2$ and $y<0.9$. The full curve corresponds to the expectation from MRST. The dashed line is a straight line fit to the measured cross sections.*

The HERA-II operation will end in 2007 according to current planning. In the present envisaged running scenario HERA will operates with an electron beam starting after the summer shutdown 2004 till the end of 2005 and will switch back to positron-proton collisions for the rest of HERA-II. In this scenario, an integrated luminosity of 0.7 fb$^{-1}$ per experiment is expected. This will be a significant gain with respect to the total data taken in HERA-I. Such a large data sample will improve even further our knowledge of the PDFs, in particular at high x and $Q^2$, the precision of which is currently statistically limited.

Indeed, precision NC cross-sections at high $Q^2$ will improve the determination of $x\widetilde{F}_3$ which is sensitive to $2u_v+d_v$ with $u_v$ and $d_v$ being the valence quark densities of u and d quarks. Similarly, precision $e^+p$ CC cross sections at high $Q^2$ will allow the d quark density to be constrained better at high x. Furthermore, from a polarization asymmetry in NC cross sections, one will determine the SF $F_2^{\gamma Z}$ which can be expressed in terms of the d/u ratio. All these data will thus hopefully clarify the controversial value of d/u at x=1 [7].

Apart from the high luminosity running at largest possible beam energies, a direct measurement of $F_L$ will need data at different CM energies. Such a measurement is important to verify the gluon density that is currently constrained mainly indirectly from the scaling violation of the SF data (Fig.1). A lower proton energy running would also help to access high x region at low $Q^2$.

Beyond HERA-II, there is strong interest [8] in the community to continue the HERA physics programme and to fully exploit its unique potential. In particular it is expected that by replacing the proton beam with a heavy nuclear beam, one can study a new QCD regime in which the gluon parton density becomes so high that saturation would take place. This however needs funding and solutions for a new injector replacing PETRA, the actual injector, which will be rebuilt to be used for XFEL (X-ray Free Electron Laser) from 2007 on.

## Acknowledgments

The author wishes to thank J. Soffer for organizing such a stimulating workshop and colleagues in H1 and ZEUS for the results shown in this contribution.